# Algorithmic Solutions to Some Transportation Optimization Problems with Applications in the Metallurgical Industry

Mugurel Ionuț Andreica*, Sorin Briciu**, Mădălina Ecaterina Andreica***

\* Politehnica University of Bucharest, Bucharest, Romania, email: mugurel.andreica@cs.pub.ro
\*\* "1 Decembrie 1918" University of Alba Iulia, Alba Iulia, Romania, email: sbriciu@yahoo.com
\*\*\* Academy of Economic Studies, Bucharest, Romania, email: madalina.andreica@gmail.com

## ABSTRACT

In this paper we address several constrained transportation optimization problems (e.g. vehicle routing, shortest Hamiltonian path), for which we present novel algorithmic solutions and extensions, considering several optimization objectives, like minimizing costs and resource usage. All the considered problems are motivated by practical situations arising, for instance, in the mining and metallurgical industry or in data communication. We restrict our attention to transportation networks with path, tree or geometric structures, for which the developed polynomial-time algorithms are optimal or nearly optimal.

## 1. INTRODUCTION

Transportation optimization is a very important topic in many domains, like mining, metallurgical and food industry, airplane and train scheduling, computer networks (file transfers, live and on-demand audio/video streaming), distribution of natural resources (natural gas, oil), and so on. Each of these fields has its own specific transportation problems, but, from a broader point of view, there are many similarities. In this paper we consider several constrained optimization problems for transportation networks with restricted topologies (e.g. networks with tree or path structures, or networks embedded in the plane). Some of these problems have applications in multiple fields (e.g. mining and metallurgical industry, as well as network data transfers), given an appropriate interpretation of the problem parameters.

The rest of this paper is structured as follows. In Section 2 we present the (relaxed) open vehicle routing problem in tree networks, for which we developed several efficient, polynomial-time algorithms. In Section 3 we consider a single vehicle routing problem with DFS constraints in trees, with the objective of minimizing the total fuel consumption. In Section 4 we present new algorithmic approaches and extensions for the well-known jeep problem. In Section 5 we present improved algorithms for computing the shortest Hamiltonian path fully contained inside a simple polygon (where the vertices of the polygon are also the graph's nodes). Finally, in Section 6 we discuss related work and in Section 7 we conclude and discuss future work.

## 2. THE (RELAXED) OPEN VEHICLE ROUTING PROBLEM IN TREES

In the (relaxed) *Open Vehicle Routing Problem*, we are given a connected, undirected graph composed of $n$ vertices and $m$ edges. Each edge $(u,v)$ has a length $l(u,v)$. A central depot (one of the graph's vertices) contains $p$ vehicles. Each vehicle $i$ $(1 \leq i \leq p)$ will be sent along a (possibly self-intersecting) path $P(i)$ in the graph. In the standard *Vehicle Routing Problem*, the path must start and end back at the central depot, but in the *open* version considered here, each vehicle's path starts at the depot, but may end in any of the graph's vertices. Each vertex in the graph must be served by at least one of the vehicles. Thus, for each graph vertex $u$, there must be a vehicle $i$, such that $u \in P(i)$. Given these constraints, we consider an objective function based on the lengths of the paths $P(i)$ of the vehicles (the length of a path is the sum of the lengths of the edges composing the path; if an edge appears $k \geq 1$ times in a path, then its length is added $k$ times). Two standard objective functions are: minimizing the sum of the lengths of the paths and minimizing the maximum length of a path (we can also consider a third case, where every vehicle $i$ has a weight $w(i)$ and we want to minimize the maximum weighted length of every path, i.e. the product between the weight of the vehicle and the length of the vehicle's path or the sum of the weighted lengths).

We will consider the objective of minimizing the total length of the vehicles' paths, in the restricted case when the network is a tree (a connected, acyclic, undirected graph). We will start with a greedy algorithm, which was brought to our attention by Prof. S. Ciurea, in a personal communication. We will root the tree at the vertex containing the central depot ($r$), thus defining parent-son relationships. Every vertex $u$, except $r$, will have a unique parent, denoted by *parent(u)*. All the neighbors of a vertex $u$, except *parent(u)*, will be its sons. We denote by $ns(u)$ the number of sons of vertex $u$ and by $s(u,1), \ldots, s(u,ns(u))$ the sons of vertex $u$. A leaf (or terminal) vertex is a vertex which has no sons. The subtree $T(u)$ of a vertex $u$ is defined recursively as the set of vertices composed of $u$ and the subtrees of vertex $u$'s sons, together with all the tree edges between these vertices.

Let's consider a vehicle's path as a sequence of vertices: $v_1=r, v_2, v_3, \ldots, v_k$, such that $v_i$ and $v_{i+1}$ ($1 \leq i \leq k-1$) are neighbors (the edge $(v_i, v_{i+1})$ exists in the tree). A vertex may appear multiple times in this sequence. It is easy to notice that, in an optimal strategy, the final vertex visited by every vehicle is a leaf vertex. Let's assume that the final vertex $v$ of a path $P(q)$ is not a leaf vertex. In this case, there exists at least one leaf $l$ in $T(v)$ and this leaf has to be visited by at least one vehicle. We can either extend the path of $P(q)$ to include $l$ or $l$ is visited by another vehicle $q'$, which also visits the vertex $v$. In this case, we can reduce the path $P(q)$ by one vertex. We can use this reasoning repeatedly, until the final vertex of every vehicle's path is a leaf.

We will color the tree vertices using two colors, *red* and *blue*, according to the following rule. A vertex $u$ is colored *red* if the edges *(parent(u), u)* and *(u, parent(u))* appear together in the path of at least one vehicle. All the other vertices are colored *blue*. Note that the tree's root is always a *blue* vertex. We will start with all the tree vertices colored *red* (except for the root) and a total cost equal to twice the sum of the costs of all the edges in the tree. This cost corresponds to a vehicle sent out from the root which traverses every edge twice and then returns to the root. Obviously, this vehicle visits all the vertices of the tree. We will repeatedly improve this solution, in a greedy fashion. Let's assume that $C_{total}$ is the current total cost of the vehicle routing strategy. At each step, we will consider all the leaves $l$ which are colored in *red*. For each such leaf $l$, we search for the vertex $v=closest\_blue(l)$ on the path from the tree root to $l$, which is colored in blue and is the closest to $l$. We now compute $\delta(l)=path\_cost(root, closest\_blue(l)) - path\_cost(closest\_blue(l), l)$, where $path\_cost(u,v)$ denotes the sum of the costs of the edges on the path between vertices $u$ and $v$; one of the two vertices must be an ancestor of the other one. At the end of each step we choose the leaf vertex $l$ for which $\delta(l)$ is minimum (and negative). If $\delta(l)$ is negative, we color all the vertices on the path between $l$ and $closest\_blue(l)$ (including the endpoints) in blue and add $\delta(l)$ to $C_{total}$. The algorithm stops when $p$ steps have been performed or when the minimum value $\delta(l)$ is not negative. In order to compute $path\_cost(u,v)$ in $O(1)$ time, we will traverse the tree and compute $droot(i)$ for each vertex $i$: $droot(r)=0$ and $droot(i) = droot(parent(i))+l(parent(i),i)$. Then, $path\_cost(u,v) = |droot(u)-droot(v)|$.

After the first step, we obtain the optimal route when only one vehicle is available. At every successive step, we introduce a new vehicle, which traverses the tree path from the root towards the leaf vertex $l$ with the minimum $\delta(l)$ value and remove a portion of the path of a previous vehicle. If we do not use all the $p$ vehicles during the algorithm, then the remaining vehicles remain inside the depot. The algorithm can easily be implemented in $O(p \cdot n^2)$ time (in each of the $O(p)$ steps we consider $O(n)$ leaves and for each leaf we search for the closest blue vertex in $O(n)$ time). We will now present a dynamic programming algorithm with an $O(p^2 \cdot n)$ time complexity. We will again root the tree at the vertex with the depot ($r$) and we will maintain the same notations as before. For each vertex $u$ of the tree, we will compute the values $C_{min}(u, P_{in}, P_{out})$=the minimum total cost of the edges traversed in $T(u)$, considering that $P_{in}$ vehicles "enter" $T(u)$ and, out of these, $P_{out}$ vehicles "leave" $T(u)$. For a leaf vertex $u$, we have $C_{min}(u, P_{in}, P_{out})=0$. For a non-leaf vertex $u$, we consider all of its sons $s(u,1), s(u,2), \ldots, s(u,ns(u))$, ordered arbitrarily. We will use an algorithm based on the principles of tree knapsack:

**ROVRP-TreeKnapsack-1(u):**
**for** $P_{in}=1$ **to** $p$ **do** { **for** $P_{out}=0$ **to** $P_{in}$ **do** { $C_{min}(u, P_{in}, P_{out})=0$ }}
**for** $j=1$ **to** $ns(u)$ **do** {
 **ROVRP-TreeKnapsack-1**$(s(u,j))$
 **for** $P_{in}=1$ **to** $p$ **do** { **for** $P_{out}=0$ **to** $P_{in}$ **do** { $C_{aux}(u, P_{in}, P_{out})=+\infty$ }}
 **for** $P_{in}=1$ **to** $p$ **do** { **for** $P_{out}=0$ **to** $P_{in}$ **do** { **for** $P'_{in}=1$ **to** $P_{in}$ **do** { **for** $P'_{out}=0$ **to** $P'_{in}$ **do** {
  **if** $(P_{out} \geq (P'_{in}-P'_{out}))$ **then** $C_{aux}(u, P_{in}, P_{out}-(P'_{in}-P'_{out})) = min\{C_{aux}(u, P_{in}, P_{out}-(P'_{in}-P'_{out})), C_{min}(u, P_{in}, P_{out}) + C_{min}(s(u,j), P'_{in}, P'_{out}) + (P'_{in}+P'_{out}) \cdot l(u,s(u,j))\}$ }}}}
 **for** $P_{in}=1$ **to** $p$ **do** { **for** $P_{out}=0$ **to** $P_{in}$ **do** { $C_{min}(u, P_{in}, P_{out})=C_{aux}(u, P_{in}, P_{out})$ }}}

The time complexity of the algorithm described above is $O(p^4 \cdot n)$. In order to improve it, we make the following observation: in an optimal strategy, the number of vehicles $P_{out}$ leaving a subtree is at most *1*. We can easily notice that, if we have $P_{out}>1$ vehicles leaving a subtree, then we can concatenate their sequences of visited vertices into one sequence and, thus, we can consider only the cases $P_{out}=0$ or *1* (of course, this decreases the number of vehicles entering the subtree, too). This is indeed possible, because the last vertex inside a subtree visited by a leaving vehicle is the root of the subtree, which is also the first vertex visited by the next leaving vehicle. With this observation, the time complexity becomes $O(p^2 \cdot n)$, but the algorithm presented above has to be modified slightly. Another observation that we do not need to use (although we could) is the following. If $P_{in}>1$ for a vertex $i$, then there is no need for any of the vehicles to leave $T(i)$. Let's assume that one of the vehicles $q$ indeed enters and leaves $T(i)$. We can obtain a solution which is not worse by not making vehicle $q$ enter $T(i)$. Instead, one of the other vehicles $v$ entering $T(i)$ may follow vehicle $q$'s path in $T(i)$. After returning back to $i$ (where vehicle $q$'s path restricted to $T(i)$ ends), vehicle $v$ may follow its normal path in $T(i)$. Thus, the only possible combinations are $(P_{in}=1, P_{out}=0$ or $1)$ and $(P_{in}>1, P_{out}=0)$. The $O(p^2 \cdot n)$ algorithm is given below:

**ROVRP-TreeKnapsack-2(u):**
for $P_{in}=1$ to $p$ do { for $P_{out}=0$ to $1$ do { $C_{min}(u, P_{in}, P_{out})=0$ }}
for $j=1$ to $ns(u)$ do {
 ROVRP-TreeKnapsack-2$(s(u,j))$
 for $P_{in}=1$ to $p$ do { for $P_{out}=0$ to $1$ do { $C_{aux}(u, P_{in}, P_{out})=+\infty$ }}
 for $P_{in}=1$ to $p$ do { for $P'_{in}=1$ to $p$ do { for $P'_{out}=0$ to $1$ do {
  if $((P'_{in}>P'_{out})$ and $(P_{in}+P'_{in}-P'_{out}-1 \leq p))$ then $C_{aux}(u, P_{in}+P'_{in}-P'_{out}-1, 0) = min\{C_{aux}(u, P_{in}+P'_{in}-P'_{out}-1, 0), C_{min}(u, P_{in}, 1) + C_{min}(s(u,j), P'_{in}, P'_{out}) + (P'_{in}+P'_{out}) \cdot l(u,s(u,j))\}$
  if $(P_{in}+1+P'_{in}-P'_{out} \leq p)$ then $C_{aux}(u, P_{in}+1+P'_{in}-P'_{out}, 1) = min\{C_{aux}(u, P_{in}+1+P'_{in}-P'_{out}, 1), C_{min}(u, P_{in}, 0) + C_{min}(s(u,j), P'_{in}, P'_{out}) + (P'_{in}+P'_{out}) \cdot l(u,s(u,j))\}$
   for $P_{out}=0$ to $1$ do { if $(P_{in}+P'_{in}-P'_{out} \leq p)$ then $C_{aux}(u, P_{in}+P'_{in}-P'_{out}, P_{out}) = min\{C_{aux}(u, P_{in}+P'_{in}-P'_{out}, P_{out}), C_{min}(u, P_{in}, P_{out}) + C_{min}(s(u,j), P'_{in}, P'_{out}) + (P'_{in}+P'_{out}) \cdot l(u,s(u,j))\}$ }}}}
 for $P_{in}=1$ to $p$ do { for $P_{out}=0$ to $1$ do { $C_{min}(u,P_{in},P_{out})=C_{aux}(u,P_{in},P_{out})$ }}}

If we consider the problem from a different angle, the time complexity can be reduced to $O(p \cdot n)$. We will consider the leaves of the tree in the DFS order $l(1), \ldots, l(k)$, where $k=O(n)$ is the total number of leaves. The DFS order implies that the leaf $l(i)$ was visited before $l(i+1)$ in a DFS traversal of the tree (starting from the root), where the sons of a vertex can be considered in any order (thus, there may be many possible DFS orders of the leaves). Any optimal solution consisting of $q \leq p$ vehicles can be modified such that each vehicle visits an interval of (consecutive) leaves from the given DFS order. We compute $C_{opt}(i, j, b)=$the minimum cost of visiting the first $i$ leaves using at most $j$ vehicles and, if $b=1$, the route of a vehicle ends at leaf $l(i)$ (if $b=0$, leaf $l(i)$ may, but needn't, be the last vertex on a vehicle's route). $C_{opt}(1, j \geq 1, *)=path\_cost(root, l(1))$ and $C_{opt}(i \geq 1, 0, *) = +\infty$. For $i>1$ and $j \geq 1$ we have: $C_{opt}(i, j, 1)=min\{C_{opt}(i-1, j, 1) + path\_cost(l(i-1), lca(l(i-1), l(i))) + path\_cost(lca(l(i-1), l(i)), l(i)), C_{opt}(i-1, j-1, 0) + path\_cost(root,l(i))\}$ and $C_{opt}(i, j, 0) = min\{C_{opt}(i, j, 1), C_{opt}(i-1, j, 0) + 2 \cdot path\_cost(lca(l(i-1), l(i)), l(i))\}$. The optimal cost is $C_{opt}(k, p, 0)$. In order to compute $lca(l(i), l(i+1))$ (the lowest common ancestor of the leaves $l(i)$ and $l(i+1)$) efficiently, we can use, for instance, the technique presented in [6]. However, if we compute and store all the $k-1=O(n)$ LCAs between two consecutive leaves $l(i)$ and $l(i+1)$ $(1 \leq i \leq k-1)$ in the beginning (by using the *level-by-level* technique), we spend only $O(n)$ time overall (in an amortized sense), because every edge of the tree is traversed at most two times. The algorithms described above compute only the optimal cost of a vehicle routing strategy, but the actual paths each vehicle follows can be determined from the values computed by the algorithms.

## 3. MINIMUM FUEL SINGLE VEHICLE ROUTING IN TREES WITH DFS CONSTRAINTS

We are given a tree with $n$ vertices. Vertex $1$ contains a depot in which a single vehicle is located. The vehicle has a tank which can store an unlimited amount of fuel. Each vertex $i$ $(1 \leq i \leq n)$ contains $g_i \geq 0$ liters of fuel, which can be collected by the vehicle and stored in its tank, at the moment the vehicle first reaches vertex $i$ (in particular, $g_1$ liters of gas can be collected immediately). Every edge $(u,v)$ has a length $l(u,v)$, meaning that the vehicle consumes $l(u,v)$ liters of fuel from its tank when traversing the edge $(u,v)$ (in any direction). We must determine a route for the vehicle, such that every vertex is visited at least once, the vehicle returns at vertex $1$ and the amount of fuel in the vehicle's tank never drops below $0$ (but can reach $0$). In order to be able to travel along a route, the vehicle needs to have an initial amount of fuel $C \geq 0$ in its tank (before collecting the $g_1$ liters of gas in vertex $1$ and starting traveling along the route). It is obvious that different routes may require a different initial amount of fuel $C$. We are interested in finding a route which minimizes $C$. However, the problem seems to be too difficult without additional constraints, because, in an optimal route, the vehicle may traverse an edge $(u,v)$ any number of times. Thus, we will consider the restricted case in which every edge must be traversed at most two times (because the vehicle must return to the depot, this restriction makes the vehicle traverse every edge *exactly* two times). These types of constraints were considered in [10] and were called *DFS constraints*. We will consider the tree rooted at vertex $1$. For each vertex $i$, we will compute $C_{min}(i)=$the minimum amount of fuel the vehicle needs to have in its tank in the beginning, if we restrict the problem to vertex $i$'s subtree, $T(i)$ (i.e. we consider that the vehicle begins its route at vertex $i$, visits every vertex in $T(i)$ at least once and traverses every edge in $T(i)$ two times). We will compute these values bottom-up (from the leaves towards the root). For a leaf vertex $i$, we have $C_{min}(i)=0$. In order to be able to compute $C_{min}(i)$ for a non-leaf vertex $i$, we will need to preprocess the tree and compute several auxiliary values.

We will first compute the values $lsum(i)$=the sum of the lengths of the edges in $T(i)$. We have $lsum(i)=0$ for a leaf-vertex and $lsum(i) = \sum_{j=1}^{ns(i)} l(i, s(i, j)) + lsum(s(i, j))$ for a non-leaf vertex $i$. We will also compute the values $gsum(i)$=the sum of the $g_j$ values of the vertices $j$ in $T(i)$. We have $gsum(i)=g_i$ plus the sum of the $gsum(s(i,j))$ values ($1 \leq j \leq ns(i)$). With these values, we can define for each vertex $i \neq 1$ the fuel profit $fprofit(i)=gsum(i)-2 \cdot lsum(i)-2 \cdot l(parent(i),i)$, representing the difference between the amount of fuel the vehicle will have after visiting all the vertices in $T(i)$ and returning to $parent(i)$, and the amount of fuel the vehicle has at vertex $parent(i)$, before starting to travel towards vertex $i$. We also define $F_{min}(i)=max\{C_{min}(i) + l(parent(i),i), -fprofit(i)\}$, the minimum amount of fuel the vehicle needs to have when it is located in the vertex $parent(i)$, in order to be able to travel towards vertex $i$, visit all the vertices in $T(i)$ and the return to vertex $parent(i)$. The values $fprofit(1)$ and $F_{min}(1)$ are not defined.

In order to compute $C_{min}(i)$ for a non-leaf vertex $i$, we will binary search this value between $0$ and $C_{max}$, where $C_{max}$ is a good upper bound (e.g. twice the sum of the lengths of all the edges in the tree). Let's assume that the binary search chose a candidate value $C_{cand}$. We need to perform a feasibility test and verify if the vehicle can visit all the vertices in $T(i)$ and return to vertex $i$ having only $C_{cand}$ liters of gas initially in the tank (and satisfying all the constraints). It is obvious that if $C_{cand}$ is a feasible value, then any value $C'>C_{cand}$ is also feasible. In order to perform the feasibility test, we will maintain a value $CFuel$, denoting the current fuel in the tank of the vehicle. We initialize $CFuel$ to $C_{cand}+g_i$. We will maintain a subset $S$ of vertex $i$'s sons which were not visited, yet (all of them, initially). We will perform the following operations at most $ns(i)$ times:

1) **select** a vertex $j \in S$, such that $CFuel \geq F_{min}(j)$ and $fprofit(j)$ is maximum among the values $fprofit(j')$ with $j' \in S$ and $CFuel \geq F_{min}(j')$.

2) **remove** vertex $j$ from $S$.

3) **set** $CFuel$ to $CFuel+fprofit(j)$

In Step *1* we choose the vertex $j$ whose subtree is visited next by the vehicle. In Step *2*, we remove vertex $j$ from the set $S$, because it is now a visited vertex. In Step *3* we update the current amount of fuel in the vehicle's tank (after visiting the vertices in $T(j)$ and returning to vertex $i$). If the set $S$ is not empty, but we cannot select any vertex in Step *1*, then the value $C_{cand}$ is not a feasible value and we will need to test a larger value; otherwise, we will test a smaller one. We can easily implement Step 1 in $O(|S|)$ time (i.e. $O(ns(i))$ time), which would lead to an $O(n^2 \cdot log(C_{max}))$ overall time complexity of the algorithm. We can improve the feasibility test by constructing a segment tree [2] over the values $F_{min}(s(i,j))$ of vertex $i$'s sons. We sort the sons such that $F_{min}(s(i,1)) \leq F_{min}(s(i,2)) \leq \ldots \leq F_{min}(s(i,ns(i)))$. We construct a segment tree with $ns(i)$ leaves. Each node of the segment tree stores a value $pmax$. Initially, for each leaf $j$ ($1 \leq j \leq ns(i)$), $pmax(leaf\ j)=fprofit(s(i,j))$. For each internal node $q$, we have $pmax(internal\ node\ q)=max\{pmax(leftson(q)), pmax(rightson(q))\}$, where $leftson(q)$ and $rightson(q)$ denote its left and right son, respectively. In order to use the segment tree at Step *1* of the feasibility test, we binary search the largest value $jmax$, such that $F_{min}(s(i,jmax)) \leq CFuel$. Then, we range query the interval $[1,jmax]$ of the segment tree and obtain the maximum $fprofit$ value, $fpmax$, of a non-visited son $s(i,j')$ with $F_{min}(s(i,j')) \leq CFuel$. If $fpmax=-\infty$, the $C_{cand}$ is not feasible, because no vertex can be selected in Step *1*. It is easy to extend the segment tree and store at each node $q$ the leaf number whose $pmax$ value is maintained at $q$. This way, we can obtain the index $j$ of vertex $i$'s son $s(i,j)$ which is selected in Step *1*. In Step *2*, we set $pmax(leaf\ j)=-\infty$ and then recompute the $pmax$ values of all the ancestors of the $j^{th}$ leaf, from the parent node of the leaf, towards the root (we set $pmax(node\ q)$ to the maximum value of the $pmax$ values of its two sons). By setting $pmax(leaf\ j)$ to $-\infty$, we virtually remove the son $s(i,j)$ from the set $S$, because its $fprofit$ value will never be selected as the maximum value again. The time complexity of the feasibility test becomes $O(ns(i) \cdot log(ns(i)))$ for a vertex $i$ and the overall time complexity of the algorithm is now $O(n \cdot log(n) \cdot log(C_{max}))$. If the $g(*)$ and $l(*,*)$ values are integers, the algorithm computes an exact solution (and the binary search finishes after the search interval becomes so small that it contains only one integer). If they are (arbitrary) real numbers, then the solution found by the algorithm is within an arbitrary constant range $\varepsilon>0$ from the optimal solution (and the binary search finishes when the length of the search interval is smaller than $\varepsilon$).

## 4. THE JEEP PROBLEM

The well-known jeep problem has been introduced and solved many years ago, by N. J. Fine, in [7]. The problem's statement is as follows. A jeep has to travel a distance of $x$ miles in the desert in a straight line and has a tank which can hold at most $m$ gallons. The gas consumption in the desert is $g$ gallons/mile. At the initial point, there is an infinite supply of gas, but no gas is available anywhere else in the desert. In order to travel through the desert, the jeep may establish several gas depots along the way, where gas is temporarily stored. The objective of the jeep is to consume the minimum amount of gas in order to travel $x$ miles.

In this section we propose a method for evaluating accurately (but slowly) the gas consumption when a given

*subdivision* (the term is defined later) is chosen; afterwards, we argue that using equal subdivisions is intuitive and useful and then we present a second method, which evaluates the gas consumption inaccurately, but fast, for equal subdivisions. Finally, we also present some interesting extensions to the jeep problem.

### 4.1. METHOD 1

Let's assume that the jeep establishes $k$ gas depots along the way, at points located at distances $d_1, d_2, ..., d_k$ from the initial point (with $d_i<d_{i+1}$ and $d_k<x$). We will add to this set the points $d_0=0$ and $d_{k+1}=x$ and call $d$ a *subdivision*. It has been shown in [7] that, if the subdivision is fixed, the optimal strategy (the one minimizing the gas consumption) consists of bringing enough gas from $d_0$ to $d_1$ (performing several round-trips between $d_0$ and $d_1$ and then one final one-way trip), then from $d_1$ to $d_2$, and so on. A function $f(i,d)$ can be computed, representing the minimum amount of gas required to reach the point $d_{k+1}$, starting from $d_i$, using the subdivision $d$. It is obvious that $f(k+1,d)=0$. For $0 \leq i \leq k$, $f(i,d)$ can be computed from $f(i+1,d)$. Let's assume that the jeep performs $rt_i$ round-trips between $d_i$ and $d_{i+1}$. For each round-trip, the jeep requires $2 \cdot g \cdot (d_{i+1}-d_i)$ gallons and, thus, can deposit at $d_{i+1}$ an amount of gas equal to $m-2 \cdot g \cdot (d_{i+1}-d_i)$. In the final trip, the jeep can deposit at $d_{i+1}$ any amount of gallons $q_i$ between $0$ and $m-g \cdot (d_{i+1}-d_i)$. Thus, the total amount of gas that is deposited at $d_{i+1}$ is $f(i+1,d) = (rt_i \cdot (m-2 \cdot g \cdot (d_{i+1}-d_i)) + q_i)$ and the total amount of gas consumed is $f(i,d) = (rt_i \cdot m + q_i + g \cdot (d_{i+1}-d_i))$. We compute $l_{i+1} = f(i+1,d) \ div \ (m-2 \cdot g \cdot (d_{i+1}-d_i))$ (where $div$ denotes integer division, i.e. the integer part of the result) and $r_{i+1} = f(i+1,d) - l_{i+1} \cdot (m-2 \cdot g \cdot (d_{i+1}-d_i))$. If $r_{i+1} \leq g \cdot (d_{i+1}-d_i)$ and $l_{i+1}>0$ then we set $rt_i=l_{i+1}-1$ and $q_i=r_{i+1}+m-2 \cdot g \cdot (d_{i+1}-d_i)$; otherwise, we set $rt_i=l_{i+1}$ and $q_i=r_{i+1}$. Using this procedure repeatedly, we can compute $f(0,d)$, which is the minimum amount of gas required to traverse $x$ miles, using the subdivision $d$.

### 4.2. USING EQUAL SUBDIVISIONS

The jeep problem requires the computation of $min\{f(0,d)\}$, over all the subdivisions $d$. In [7], the author presented the subdivision which minimizes the gas consumption and also made another interesting observation. If we use an *equal subdivision* $d$ with sufficiently many points, then $f(0,d)$ can get arbitrarily close to the minimum amount of gas required. An equal subdivision $d_0=0, d_1, ..., d_k, d_{k+1}=x$ has the property that $d_{i+1}-d_i=c=x/(k+1)$. Using an equal subdivision in order to try to solve the problem is definitely more intuitive than using the subdivision provided by Fine in [7]. Furthermore, there are situations when we are interested in obtaining a gas consumption which is lower than a given threshold and not necessarily minimum. For such situations, we can start with an equal subdivision $d^{(1)}$ having $k_1$ points and compute the value $f(0,d^{(1)})$ using the method described previously (method *1*). If $f(0,d^{(1)})$ is too large, we can increase the number of points to $k_2>k_1$ (e.g., we can set $k_2=ct \cdot k_1$ or $k_2=k_1+ct$, where $ct$ denotes a constant) and, consequently, use a different equal subdivision $d^{(2)}$. Then, we compute $f(0,d^{(2)})$. We can repeat this process until we obtain a value which is small enough. The problem posed by this method is that it takes $O(k)$ time for computing $f(0,d)$ for a subdivision with $k$ points. As the number of points of the subdivision increases, this method becomes slower and slower. We will present here a method which is faster and makes use of the fact that we consider only equal subdivisions.

### 4.3. METHOD 2

From method *1* (which is inspired from the observations made in [7]), we have that if $f(i+1,d) = (l_{i+1} \cdot (m-2 \cdot g \cdot (d_{i+1}-d_i)) + r_{i+1})$ and $r_{i+1}>g \cdot (d_{i+1}-d_i)$, then $f(i,d) = (l_{i+1} \cdot m + r_{i+1} + g \cdot (d_{i+1}-d_i)) = (f(i+1,d) + l_{i+1} \cdot 2 \cdot g \cdot (d_{i+1}-d_i) + g \cdot (d_{i+1}-d_i))$. We will first change the method slightly and ignore the case $r_{i+1} \leq g \cdot (d_{i+1}-d_i)$ (and $l_{i+1}>0$). Thus, we will change method *1* such that we always set $rt_i=l_{i+1}$ and $q_i=r_{i+1}$. We will consider that this method computes the values $g(i,d)$ ($g(k+1,d)=0$). Let's assume that $g(i+1,d)=f(i+1,d)$ and the case $r_{i+1} \leq g \cdot (d_{i+1}-d_i)$ (and $l_{i+1}>0$) occurs. The correct value $f(i,d)$ is $((rt_i-1) \cdot m + r_{i+1} + m-2 \cdot g \cdot (d_{i+1}-d_i) + g \cdot (d_{i+1}-d_i))$, whereas the computed value $g(i,d)$ is $(rt_i \cdot m + r_{i+1} + g \cdot (d_{i+1}-d_i))$. The difference $g(i,d)-f(i,d)$ is equal to $2 \cdot g \cdot (d_{i+1}-d_i)$. Once a value $g(i,d)$ is larger than $f(i,d)$, all the values $g(j,d)$, with $j<i$, will be larger than the corresponding $f(j,d)$ values. By changing the method, the computed values are no longer correct. However, by ignoring the case $r_{i+1} \leq g \cdot (d_{i+1}-d_i)$ (and $l_{i+1}>0$), this second method can be made to run faster. We should mention that $g(0,d)$ is at least as large as $f(0,d)$ and, thus, whenever $g(0,d)$ is found to be satisfactory (in the context presented in subsection B), $f(0,d)$ will also be satisfactory. We believe that the loss of accuracy of the method is compensated by the increase in speed. In this second method, we always have $g(i,d) = (g(i+1,d) + l_{i+1} \cdot 2 \cdot g \cdot (d_{i+1}-d_i) + g \cdot (d_{i+1}-d_i))$, where $l_{i+1} = (g(i+1,d) \ div \ (m-2 \cdot g \cdot (d_{i+1}-d_i)))$. For an equal subdivision $d$, we have $d_{i+1}-d_i=c$ and we will denote the constant quantity $g \cdot c$ by $a$. Thus, $g(i,d)=(g(i+1,d) + l_{i+1} \cdot 2 \cdot a + a)$, where $l_{i+1} = (g(i+1,d) \ div \ (m-2 \cdot a))$. We make the following observation: if all the values $g(j,d)$, with $u \leq j \leq v$, have the property that $g(j,d) \ div \ (m-2 \cdot a)=l(u,v)$ (the same value $l(u,v)$) and we know the value $g(v,d)$, then we can easily compute the value $g(u-1,d)$, without computing the intermediate values $g(u,d), g(u+1,d), ..., g(v-1,d)$: $g(u-1,d) = (g(v,d) + (v-u+1) \cdot (l(u,v) \cdot 2 \cdot a+a))$. Furthermore, assuming that we know the value $g(v,d)$ (and, consequently, the value $l_v$), we can easily compute the smallest index $u=first(v)$, such that $g(j,d) \ div \ (m-2 \cdot a)=l_v=l(u,v)$, for all $u \leq j \leq v$. A simple approach would be to binary search $u$ between $1$ and $v$, compute the potential value $g'(u,d) =$

$(g(v,d) + (v-u) \cdot (l_v \cdot 2 \cdot a+a))$ and verify if we obtain $l_v$ when we divide it by $(m-2 \cdot a)$. If we obtain a result which is larger than $l_v$, then we will test a larger value of $u$ next; otherwise, we will consider a smaller value. We can also compute $u$ directly. First, we compute $r_v = g(v,d) - l_v \cdot (m-2 \cdot a)$. Then, we divide $(m-2 \cdot a - r_v)$ by $(l_v \cdot 2 \cdot a+a)$ and obtain an integer number $dif_v = ((m-2 \cdot a - r_v) \text{ div } (l_v \cdot 2 \cdot a+a))$. $(m-2 \cdot a - r_v)$ is the maximum amount by which the value $g(v,d)$ can be increased such that when dividing it by $(m-2 \cdot a)$ we still obtain the quotient $l_v$. Since a value $g(i,d)$ increases by $l_{i+1} \cdot 2 \cdot a+a$ from the value $g(i+1, d)$, it is natural to divide $(m-2 \cdot a-r_v)$ by $(l_v \cdot 2 \cdot a+a)$, in order to find out how many values before $v$ give the same quotient $l_v$ when divided by $(m-2 \cdot a)$; if $(m-2 \cdot a-r_v)$ is an integer multiple of $(l_v \cdot 2 \cdot a+a)$, then we will decrease $dif_v$ by 1 The required value $u=first(v)$ is $max\{v-dif_v,1\}$.

The method is now quite simple. Considering an equal subdivision $d_0=0, d_1, ..., d_{k+1}=x$, we start from $idx=k+1$ and $g(k+1,d)=0$. Then, while $(idx>0)$, we perform the following three steps: *i)* compute $first(idx)$; *ii)* compute $g(first(idx)-1, d)$; *iii)* set $idx=first(idx)-1$. Using this method, we skip over a large number of points from the subdivision. In fact, we consider at most $min\{l_0+2,k+2\}$ points, where $l_0=g(0,d) \text{ div } (m-2 \cdot a)$. There are, however, two concerns that we haven't satisfactorily addressed, yet. The first one refers to estimating the ratio $g(0,d)/f(0,d)$, for an equal subdivision $d$, preferably in terms of the number of points of the subdivision. The second concern refers to the actual number of values computed by method *2*. Again, it would be nice to estimate this number in terms of the number of points of a subdivision. So far, we have performed only some computational experiments, which confirmed the following expectations: as the number of points of an equal subdivision $d$ increases (and, thus, the distance between the points decreases), $g(0,d)$ tends towards $f(0,d)$. Regarding the number of values computed by the method *2* compared to those computed by method *1*, we computed the running times of the *2* methods on equal subdivisions. Let's assume that $R_i(k)$ $(i=1,2)$ is the running time of method $i$ on an equal subdivision with $k+2$ points. We noticed that $R_2(k)/R_1(k)$ decreases as $k$ increases.

## 4.4. PROBLEM EXTENSIONS

In this section we would like to propose some interesting extensions to the Jeep problem, which, as far as we are aware, have not been considered before. We can model the desert as an undirected graph with $n$ vertices, in which the jeep leaves from vertex *1* and wants to reach vertex *n* with a minimum amount of gas consumed. Each edge *(i,j)* has a length *len(i,j)*. Gas can be found only at vertex *1* and gas depots can be established only at the graph vertices. In order to solve this extension, we can use Dijkstra's algorithm, in order to compute the values $h(i)$=the minimum amount of gas consumed in order to reach vertex *n* (starting from vertex *i*). Initially, we have $h(n)=0$ and $h(i)=+\infty$, for $1 \leq i \leq n-1$. When expanding a vertex *i* in Dijkstra's algorithm, we consider all of its neighbors *j*. Then, using the first method presented in this note, we can consider vertices *j* and *i* as part of a subdivision, in which vertex *j* precedes vertex *i* and the distance between them is *len(i,j)*. Thus, we can compute a candidate value *cand(i,j)* and compare it to the current value *h(j)*. If *cand(i,j)<h(j)*, we set *h(j)=cand(i,j)* and update the corresponding data structures (e.g. the min-heap, if we use a priority queue in the implementation of Dijkstra's algorithm). We can also solve this problem by binary searching the minimum amount of gas $G_{min}$ available at vertex *1*. Then, as a feasibility test, we compute *hmax(i)*=the maximum amount of gas which can be brought from vertex *1* to vertex *i* (where $hmax(1)=G_{min}$). We can use Dijkstra's algorithm again, but in a forward manner this time. It is easy to compute a candidate value *cand(i,j)* when expanding a vertex *i* which is a neighbor of vertex *j*. If $2 \cdot g \cdot len(i,j) \geq min\{hmax(i), m\}$, then $cand(i,j) = min\{hmax(i),m\}-g \cdot len(i,j)$. Otherwise, we consider $hmax(i)=q(i) \cdot m+r(i)$, where $q(i)$ is the integer quotient of dividing $hmax(i)$ by $m$. If $r(i)<g \cdot len(i,j)$, then $cand(i,j)=c_1(i,j)=(q(i)-1) \cdot (m-2 \cdot g \cdot len(i,j))+m-g \cdot len(i,j)$. Otherwise, $cand(i,j) = max\{c_1(i,j), q(i) \cdot (m-2 \cdot g \cdot len(i,j)) + r(i)-g \cdot len(i,j)\}$. Another extension is similar to the previous one, except that gas depots can be established anywhere (both at graph vertices and along the edges). In this case, we compute the shortest path from vertex *1* to vertex *n*. Let *x* be the length of this path. We now use the method described in [7], as if the jeep had to travel *x* miles into the desert.

A third extension allows the jeep to establish gas depots anywhere along the edges, but forces it to establish gas depots at the graph vertices. In this case, we use Dijkstra's algorithm again and compute the same values $h(i)$ as in the first extension. When expanding a vertex *i* and considering a neighboring vertex *j*, we compute a candidate value *cand(i,j)*=the minimum amount of gas required at vertex *j*, considering that the jeep has to travel a distance equal to *len(j,i)* (from *j* to *i*), must deliver a quantity equal to *h(i)* at vertex *i* and can establish gas depots anywhere along the way. For this, we can use the results in [20] or the first or second methods described in this section together with an equal subdivision $d_0=0, d_1, ..., d_k, d_{k+1}=len(i,j)$ with a sufficiently large number of points, where we have $g(k+1,d)=h(i)$, instead of *0*. If *cand(i,j)<h(j)*, then we set *h(j)=cand(i,j)*.

## 5. SHORTEST HAMILTONIAN PATH INSIDE A SIMPLE POLYGON

We are given a simple polygon with *n* vertices. We want to find a shortest Hamiltonian path inside the polygon, i.e. a path of minimum length which starts at some vertex of the polygon and visits every other vertex exactly once. We will consider two variations of this problem, when the starting vertex is given and when it is not, and we will present dynamic programming algorithms for both situations. The applications of this problem are obvious. The polygon vertices may be guard posts and the area inside the polygon may be the only safe area. The starting vertex is a supply center and a vehicle must transport supplies to every other vertex of the polygon taking the least amount of time, without leaving the safe area (i.e. the polygon). The shortest Hamiltonian path problem inside a simple polygon with $n$ vertices, when the starting vertex is given, was considered in [5] and solved in $O(n^3)$ time, using dynamic programming. We will present here an $O(n^2)$ solution, inspired from the ideas introduced in [5]. Let's consider the polygon's vertices numbered from $0$ to $n-1$ and let's consider vertex $0$ as the starting vertex. The algorithm computes two tables, $A(i,j)$ and $B(i,j)$. $A(i,j)$ is the minimum length of a path which contains every vertex in the interval $[i,j]$ (which also contains vertex $0$) and ends at vertex $i$. $B(i,j)$ has the same meaning, except that the path ends at vertex $j$. An interval $[a,b]$ is defined as the set $\{a, (a+1) \bmod n, …, b\}$ (i.e. it is considered circularly along the contour of the polygon).

Before computing these values, we need to compute the visibility graph between the vertices of the polygon. Two vertices $i$ and $j$ are visible if the straight line segment which connects them is fully contained inside the polygon. This can be achieved in $O(n^2)$ time [22]. If the polygon is convex, then any pair of vertices is visible and we do not need to compute any visibility information. We have $A(0,0)=B(0,0)=0$ and $A(i,i)=B(i,i)=+\infty$ $(i>0)$. We will compute the values in increasing order of the number of vertices contained in the interval $(i,j)$ (this number is $((n-i) \bmod n)+j+1$ if $i>j$, and $(j-i+1)$ if $i \leq j$). For any interval $[i,j]$ which does not contain the vertex $0$, we have $A(i,j)=B(i,j)=+\infty$. When $[i,j]$ contains vertex $0$, we have: $A(i,j) = min\{A((i+1) \bmod n, j) + dist((i+1) \bmod n, i), B((i+1) \bmod n, j) + dist(j,i)\}$ and $B(i,j) = min\{B(i,(j-1+n) \bmod n) + dist((j-1+n) \bmod n, j), A(i, (j-1+n) \bmod n) + dist(i,j)\}$. The function $dist(p,q)$ returns (in $O(1)$ time, obviously) the distance between the vertices $p$ and $q$ of the polygon, if they are visible, or $+\infty$ otherwise. By using the $dist(*,*)$ function we can also introduce different scenarios, like weighted distances; we can even set $dist(i,j)=+\infty$ whenever we want to disallow the direct connection between some pairs of vertices $(i,j)$. However, in other scenarios, we must add the condition that no two segments connecting consecutive vertices along the path cross. The length of the shortest Hamiltonian path can be found at the entry $A((j+1) \bmod n, j)$ or $B((j+1) \bmod n, j)$ whose value is minimum. In order to compute the actual path, we can easily trace back the way the $A(*,*)$ and $B(*,*)$ values were computed. It is obvious that every entry $A(i,j)$ or $B(i,j)$ can be computed in $O(1)$ time and, thus, the total time complexity of the algorithm is $O(n^2)$.

In order to find the shortest Hamiltonian path without a given starting vertex, the author of [5] suggests considering every vertex of the polygon as a starting vertex and obtains an $O(n^4)$ solution. Using the algorithm we presented in the previous paragraph we can obtain immediately an $O(n^3)$ solution. However, we can solve this case in $O(n^2)$, too. We will now initialize every value $A(i,i)$ and $B(i,i)$ to $0$ $(0 \leq i \leq n-1)$ and compute the other values $A(i,j)$ and $B(i,j)$ using the equations form the previous paragraph. In this case, however, there is no restriction imposed on the interval $[i,j]$ (it does not have to contain the vertex $0$). A description of the algorithm is given below:

**ShortestHamiltonianPath():**
**for** $i=0$ **to** $n-1$ **do** { $A(i,i)=B(i,i)=0$ }
**for** $k=1$ **to** $n-1$ **do** { **for** $i=0$ **to** $n-1$ **do** {
 $j=(i+k) \bmod n$
 $A(i,j) = min\{A((i+1) \bmod n, j) + dist((i+1) \bmod n, i), B((i+1) \bmod n, j) + dist(j,i)\}$
 $B(i,j) = min\{B(i, (j-1+n) \bmod n) + dist((j-1+n) \bmod n, j), A(i, (j-1+n) \bmod n) + dist(i,j)\}$ }}

The length of the shortest Hamiltonian path is the minimum of the entries $A((j+1) \bmod n, j)$ or $B((j+1) \bmod n, j)$ and the path can be computed by tracing back the way we computed the $A(*,*)$ and $B(*,*)$ values. We can also consider other objectives besides the length of the Hamiltonian path, but only if we further restrict the problem. Let's assume that the $n$ vertices are located on a closed, non-intersecting curve and the vehicle can only travel along the curve. Let's assume that we know the distances $d_i$ between every two consecutive vertices along the curve, $i$ and $(i+1) \bmod n$. In this case, the shortest Hamiltonian path is trivial. If the starting vertex is $0$, then the vehicle needs to travel along the curve in one direction or the other and the length is the total length of the curve, minus $max\{d_0, d_{n-1}\}$. When the starting vertex is not given, we will choose it in such a way that the maximum distance between two consecutive points is not part of the vehicle's path. Thus, the minimum length is equal to the total length of the curve, minus $max\{d_i | 0 \leq i \leq n-1\}$.

Let's consider now that every vertex $i$ has a weight $w_i$ and we want to find a Hamiltonian path which minimizes the sum of weighted distances from the starting vertex to every other vertex along the path. To be more precise, we define $dt(i)$=the total distance traveled by the vehicle before first reaching vertex $i$. We want to minimize the sum of the $w_i \cdot dt(i)$ values (with $0 \leq i \leq n-1$). We will use dynamic programming again and we will compute the same values as before: $A(i,j)$= the minimum sum of weighted distances, considering that the vehicle visited every vertex in the interval $[i,j]$ and is now located at vertex $i$; $B(i,j)$ is defined similarly, except that the vehicle is located at vertex $j$. If the starting vertex $s$ is given, we have $A(s,s) = B(s,s) = 0$ and $A(i,i)=+\infty$ for $i \neq s$. If the starting vertex is not given, we initialize all the values $A(i,i)$ and $B(i,i)$ to $0$. We will now compute the values $A(i,j)$ and $B(i,j)$ in increasing order of the length of the interval $[i,j]$ (just like before). We have: $A(i,j) = min\{A((i+1) \bmod n, j) + dist((i+1) \bmod n, i) \cdot wsum((j+1) \bmod n, i), B((i+1) \bmod n, j) + dist(j,i) \cdot wsum((j+1) \bmod n, i)\}$ and $B(i,j)=min\{A(i, (j-1+n) \bmod n) + dist(i,j) \cdot wsum(j, (i-1+n) \bmod n), B(i, (j-1+n) \bmod n) + dist((j-1+n) \bmod n, j) \cdot wsum(j, (i-1+n) \bmod n\}$. We denote by $wsum(a,b)$ the sum of the weights of the vertices in the interval $[a,b]$. In the previous equations, $wsum(a,b)$ denotes the sum of the weights of the vertices outside the interval of already visited vertices; this sum is then multiplied by the distance traveled during the most recent step. We can compute this value in $O(1)$ time, if we first preprocess the weights of the vertices and compute an array $wp(i)$=the sum of the weights of the vertices in the interval $[0,i]$. We have $wp(-1)=0$ and $wp(i \geq 0)=wp(i-1)+w_i$. If $(a<b)$ then $wsum(a,b) = wp(b)-wp(a-1)$; otherwise, $wsum(a,b) = wp(n-1)-wp(a-1)+wp(b)$. We denote by $dsum(i,j)$ the distance between vertex $i$ and vertex $j$ along the curve, passing only through parts of the curve and other vertices which are included in the interval $[i,j]$. With this definition, $dsum(i,j)$ is, in fact, the sum of the values $d_k$, with $k$ in the interval $[i, (j-1+n) \bmod n]$, and can be computed in $O(1)$ time if we use a similar preprocessing technique as in the case of $wsum$. Then, $dist(a,b) = min\{dsum(a,b), dsum(b,a)\}$. It should now be obvious that we can compute every value $A(i,j)$ and $B(i,j)$ in $O(1)$ time, obtaining an $O(n^2)$ overall time complexity.

## 6. RELATED WORK

Transportation problems are of major importance because of their immediate applications in real-life situations. The vehicle scheduling problem is a more general version of the relaxed vehicle routing problem considered in Section 2. Each vertex of the graph can only be served during a given time interval and serving it takes a certain amount of time. The complexity of the *Single Vehicle Scheduling Problem* on several classes of graphs was studied in [18]. The same problem, but on tree networks, was studied in [10]. The *Vehicle Routing Problem* is slightly different than the *Vehicle Scheduling Problem*. Every vertex must be visited by exactly one vehicle and the vehicles must return to the central depot at the end of their route. Routes are subject to constraints, like total path length or maximum path length constraints. In [12], the authors present a survey of exact and approximate algorithms for the *Vehicle Routing Problem*. Many variations of the *Vehicle Routing Problem* were studied, e.g. the *Minimum Vehicle Routing* with a common deadline [17], the *Very Offline k-Vehicle Routing Problem in Trees* [16] and the *Preemptive Vehicle Routing Problem in Trees* [8]. The *Open Vehicle Routing Problem* (in which the vehicles do not need to return to the depot) was studied under several variations in [15]. The *Traveling Salesman Problem*, together with its many variations [13, 19], is strongly related to some of the problems considered in this paper (particularly those from Sections 2, 3 and 5). The shortest Hamiltonian problem inside a simple polygon was previously considered in [5], where $O(n^3)$ ($O(n^4)$) algorithms were given for the case when the starting vertex is (is not) given. Maintaining the core ideas presented in [5], we were able to improve the algorithms for both cases to $O(n^2)$. The *Jeep Problem*, as defined in Section 4, was analyzed in [7] and [20]. Several extensions, variations and different solution approaches have been considered and proposed in [9, 11]. In many real-life transportation optimization problems we need to also be able to cope with uncertainty; we can do this by using special techniques, e.g. fuzzy sets [21], or by explicitly introducing probabilities as problem parameters (see [1] for an approach based on risk analysis, or [3] and [4] for some reliability models for networks with tree topologies).

## 7. CONCLUSIONS AND FUTURE WORK

In this paper we considered several transportation and vehicle scheduling optimization problems, like: 1) open vehicle routing in trees subject to minimizing the total length of the paths of the vehicles, 2) single vehicle routing in trees with DFS constraints, subject to minimizing the initial amount of fuel, 3) the well-known jeep problem and some of its extensions, and 4) the shortest Hamiltonian path inside a simple polygon. For each problem we presented efficient, novel, optimal or nearly optimal algorithms, which bring significant improvements upon the previously known solutions to these problems or to similar ones. All the considered problems are motivated by real-life situations occurring in the mining and metallurgical industry, where the transportation of the required materials is a very important issue. We also tested the efficiency of some of the developed algorithms to some concrete situations from the metallurgical industry and we found the results to be promising. As future work, we intend to extend the developed algorithms to networks with a more general structure, as well as consider more complex cost calculation methods [14].